\documentclass[10pt]{iopart}

\usepackage{graphicx}
\usepackage{epsfig}
\usepackage{xcolor}
\usepackage{iopams}
\usepackage[latin1]{inputenc}

\begin{document}

\title[ESR of the itinerant ferromagnets RECrGe$_{3}$]{Electron Spin Resonance of the itinerant ferromagnets LaCrGe$_{3}$, CeCrGe$_{3}$ and PrCrGe$_{3}$}

\author{J\"org Sichelschmidt$^1$, Thomas Gruner$^1$\footnote{Present address: Cavendish Laboratory, University of Cambridge, Cambridge CB3 0HE, United Kingdom}}

\address{$^1$Max Planck Institute for Chemical Physics of Solids, D-01187 Dresden, Germany}

\author{Debarchan Das$^2$\footnote{Present address: Laboratory for Muon Spin Spectroscopy, Paul Scherrer Institut, CH-5232 Villigen PSI, Switzerland}, Zakir Hossain$^{2,3}$}

\address{$^2$Department of Physics, Indian Institute of Technology, Kanpur 208016, India}

\address{$^3$Institute of Low Temperature and Structure Research,
Okólna 2, 50-422 Wroclaw, Poland}
\ead{Sichelschmidt@cpfs.mpg.de}

\begin{abstract}
We report Electron Spin Resonance of the itinerant ferromagnets LaCrGe$_{3}$, CeCrGe$_{3}$, and PrCrGe$_{3}$. These compounds show well defined and very similar spectra of itinerant Cr 3$d$ spins in the paramagnetic temperature region.  
Upon cooling and crossing the Cr-ferromagnetic ordering (below around 90~K) strong spectral structures start to dominate the resonance spectra in a quite different manner in the three compounds. In the Ce- and Pr-compounds the resonance is only visible in the paramagnetic region whereas in the La-compound the resonance can be followed far below the ferromagnetic ordering temperature. This behavior will be discussed in terms of the specific interplay between the 4$f$ and 3$d$ magnetism which appears quite remarkable since CeCrGe$_{3}$ displays heavy fermion behavior even in the magnetically ordered state. 
\end{abstract}


\maketitle

\section{Introduction}

The interplay between local 4$f$ magnetism and itinerant 3$d$ ferromagnetism is an important topic in heavy fermion systems and superconducting pnictide systems \cite{krellner09c,sarkar10a}. This is also relevant for systems such as CeCrGe$_{3}$ which show 4$f$ Kondo lattice and heavy fermion behavior in the presence of ferromagnetic order of itinerant 3$d$ moments \cite{das14a,das16a}. A comparison of CeCrGe$_{3}$ with LaCrGe$_{3}$ revealed a remarkably large Sommerfeld coefficient \cite{das14a} which characterizes the specific heat in the magnetically ordered state. The roles of spin fluctuations and quantum criticality in itinerant ferromagnetic systems were investigated in the systems LaV$_{\rm x}$Cr$_{\rm 1-x}$Ge$_{3}$ and 
CeCr$_{\rm 1-x}$Ti$_{\rm x}$Ge$_{3}$ where a strong suppression of ferromagnetic order was observed \cite{lin13a,das15a}.
Recently, LaCrGe$_{3}$ has drawn considerable attention in the search for a pressure driven ferromagnetic quantum critical point \cite{rana21a,kaluarachchi17a,taufour16a}.
For the investigation of the spin dynamics of metals with strong ferromagnetic correlations the spectroscopic method of Electron Spin Resonance (ESR) provided important results in 3$d$ metals like TiBe$_{2}$ \cite{shaltiel87a}, ZrZn$_{2}$ \cite{walsh70a,forster10a}, NbFe$_{2}$ \cite{rauch15a} as well as 4$f$ heavy fermion metals like YbRh$_{2}$Si$_{2}$ \cite{sichelschmidt03a} and CeRuPO \cite{krellner08a,forster10b}. 
ESR is probing the magnetic ion locally and directly in contrast to magnetization measurements which probe the bulk of the sample.
Here we report the Cr$^{3+}$ ESR in LaCrGe$_{3}$, CeCrGe$_{3}$, and PrCrGe$_{3}$ in which the magnetism is based on different interactions between 3$d$  and 4$f$ electrons, namely pure itinerant 3$d$ magnetism (LaCrGe$_{3}$, 4$f^{0}$), heavy fermion ferromagnetism with 3$d$--4$f$ hybridization (CeCrGe$_{3}$, 4$f^{1}$), and itinerant 3$d$ / local 4$f$ magnetism (PrCrGe$_{3}$, 4$f^{2}$). Distinct differences in the ESR properties of these compounds are mainly found in the ferromagnetically ordered state.

\section{Experimental Details}

Polycrystalline samples of RECrGe$_{3}$ with RE = La, Ce and Pr were prepared and characterized as described previously \cite{das14a}. 
%
%
All magnetic properties were measured using a Quantum Design superconducting quantum interference device - vibrating-sample magnetometer (QD SQUID VSM). Specific heat and electrical four-point resistivity were measured in a commercial QD Physical Property Measurement System (PPMS) equipped with a $^3$He option.
The ESR experiments were carried out using a standard continuous wave spectrometer at X-band frequencies ($\nu= 9.4$~GHz). The temperature was varied between 4 and 295~K with a He-flow cryostat.

ESR probes the absorbed power $P$ of a transversal magnetic microwave field as a function of a static and external magnetic field $\mu_0H$ \cite{abragam70a}. To improve the signal-to-noise ratio, we used a lock-in technique by modulating the static field, which yields the derivative of the resonance signal $dP/dH$. The measured ESR spectra were fitted with a Lorentzian function including the influence of the counter-rotating component of the linearly polarized microwave field \cite{rauch15a}. From the fit, we obtained the resonance field $H_{res}$ (which determines the ESR $g$-factor $g=h\nu/\mu_BH_{res}$), and the linewidth $\Delta H$ (half-width at half maximum). In metals, $\Delta H$ is a direct measure of the spin lattice relaxation time $1/T_1$ and its temperature and frequency dependences reveal the nature of the participating relaxation mechanisms.

\section{Experimental Results}

\subsection{Magnetic, transport, and thermal properties}
\begin{figure}[b]
 \centering
\includegraphics[width=0.55\textwidth]{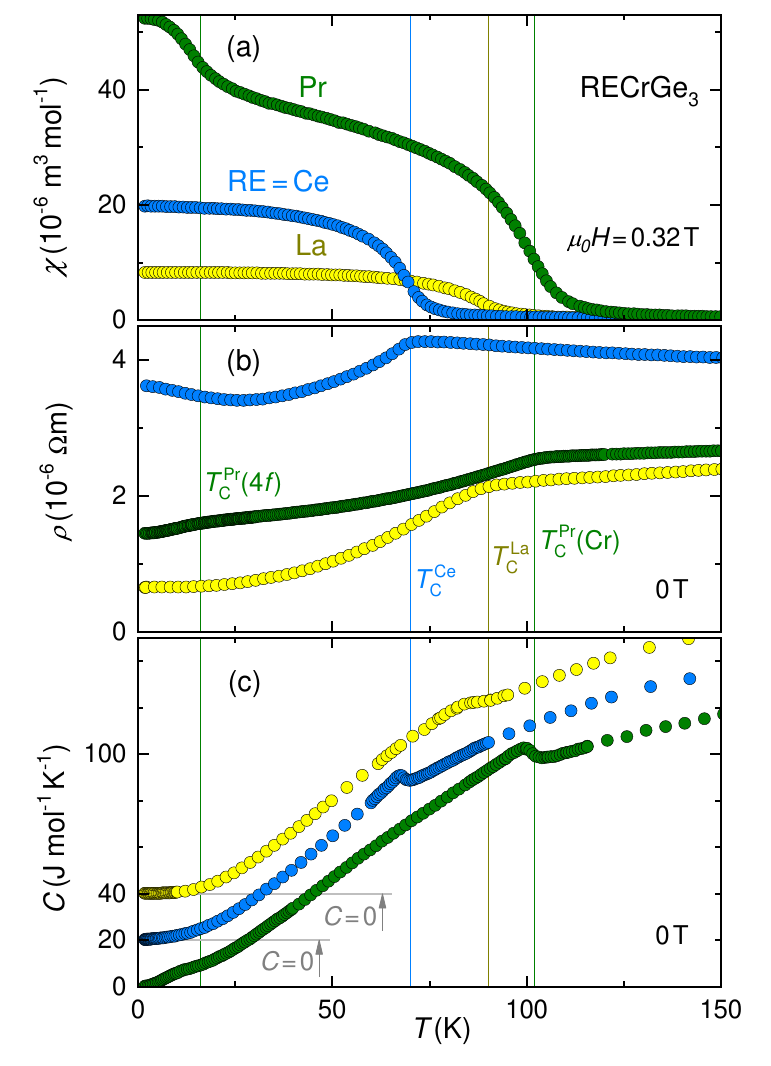}
 
\caption{Temperature dependence of (a) magnetic susceptibility $\chi(T)$, (b) electrical resistivity $\rho(T)$, and (c) specific heat $C(T)$ for LaCrGe$_{3}$, CeCrGe$_{3}$ and PrCrGe$_{3}$. At high temperatures ferromagnetic transitions are indicated by $T_{\rm C}^{\rm La, Ce, Pr}$ which are defined as the point of inflection of $\chi(T)$. At $T_{\rm C}^{\rm Pr}(4f) \approx 16$~K the Pr-compound undergoes another transition. Both $\rho(T)$ and $C(T)$ show distinct anomalies at the respective ordering temperatures, too. The $C(T)$ data of CeCrGe$_{3}$ and LaCrGe$_{3}$ are shifted by 20 and 40 $\rm{J\,mol^{-1}K^{-1}}$, respectively.} 
\label{props}

\end{figure}
\begin{figure}[bt]
 \centering
\includegraphics[width=0.55\textwidth]{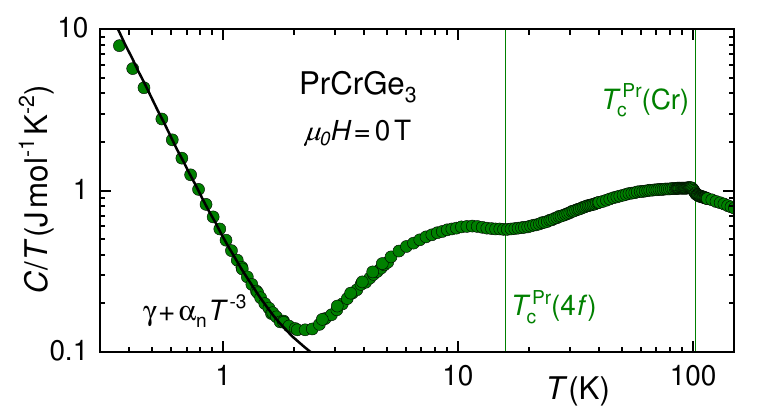}
\caption{Temperature dependence of $C/T$ of PrCrGe$_3$. At $T<2\,$K the main contribution is the high temperature tail of the nuclear Schottky peak. The solid line is a fit with $C_n/T = \gamma + \alpha_n T^{-3}$.}
\label{f2a}
\end{figure}

Measurements of the electrical transport and thermodynamic properties were carried out for PrCrGe$_3$ as was done previously for LaCrGe$_3$ and CeCrGe$_3$ in Ref. \cite{das14a}. Furthermore, we performed DC-susceptibility measurements on all RECrGe$_3$ (see figure \ref{props}a) in a field of 0.32\,T to get the magnetic response close to the ESR resonance field $H_{\rm res}$ (see figure  \ref{ESRparam}). The magnetic properties of LaCrGe$_3$ and CeCrGe$_3$ are consistent with earlier reports \cite{das14a,synoradzki19a,yang21a}.

\begin{table}[tbh]
  \caption{\label{table1}Summary of magnetic susceptibility data analysis of RECrGe$_3$.}
  \begin{indented}
  \item[]\begin{tabular}{lccccccc}
    \br
   RE & $\mu_{\rm eff, total}$ & $\mu_{\rm eff}$(RE) & $\mu_{\rm eff} (\rm Cr) $ & $T_{\rm C}^{\,\chi}(\rm Cr)$ &  $T_{\rm C}^{\,\chi}({4f})$ & $\Theta$  \\
   & $(\mu_{\rm B})$ & $(\mu_{\rm B})$ & $(\mu_{\rm B})$ & (K) & (K) & (K) \\
  \mr
  La  &2.44& 0 & 2.44 & 90  &  -   &  104  \\
  Ce &3.38& 2.54 &  2.2 & 70  &  -   &  72    \\
  Pr  &4.10& 3.58 & 2.0 & 102 &  16 &  105  \\
  \br
  \end{tabular}
  \end{indented}
 \end{table}
Interestingly, PrCrGe$_{3}$ exhibits \textit{two} anomalies, one at $\sim102$~K due to the ordering of itinerant Cr- moments, and another at $\sim16$~K suggesting a low-temperature transition due to the ordering of Pr$^{3+}$ moments. From linear fitting the inverse susceptibility data with a Curie-Weiss law ($\chi^{-1} = (T - \Theta) / C_{\rm CW}$) for the temperature range between 200~K and 400K which is well above the ordering temperature $T_{\rm C}$ (defined as the inflection point of $\chi(T)$) we obtained the effective moments ($\mu_{\rm eff}$) and Weiss temperatures ($\Theta$), see summarized values in table \ref{table1}. The Cr contribution in the effective moment $\mu_{\rm eff}$ (Cr) for CeCrGe$_3$ and PrCrGe$_3$ can be estimated assuming stable 3+ valence states for the rare earth ions. A similar approach was discussed previously \cite{das14a,das15a}. The resulting moments $\mu_{\rm eff}$(Cr) summarized in table \ref{table1} are less than expected for free Cr$^{3+}$ ions ($3.8\,\mu_{\rm B}$). These reduced moments hint to an itinerant character of the Cr magnetism in all measured RECrGe$_3$  compounds.

Positive and large values of $\Theta$ for all cases are indicative of a strong ferromagnetic exchange interaction in these systems. 
The electrical resistivity data, $\rho(T)$, in figure \ref{props}b exhibit a sharp drop below the magnetic transition temperatures due to the reduction of spin disorder resistivity. Furthermore, while LaCrGe$_3$ and PrCrGe$_3$ show metallic behavior, $\rho(T)$ for CeCrGe$_3$ is similar to that of heavy fermion materials \cite{das14a}. The resistivity anomaly at 102~K in PrCrGe$_3$ is due to magnetic order of Cr-moments and the anomaly near 16K  is likely due to the ordering of Pr$^{3+}$ moments. The kink temperatures in $\rho(T)$ are consistent with the $T_{\rm C}$ values determined from $\chi(T)$ (see figures \ref{props}a and b).

Specific heat $C(T)$ data are presented in figures \ref{props}c and \ref{f2a}. The bulk nature of the magnetic transitions are confirmed by pronounced anomalies in $C(T)$ near $T_{\rm C}$. CeCrGe$_3$ exhibits a significant enhancement of the Sommerfeld coefficient $\gamma = 130\,{\rm mJ\,mol^{-1}K^{-2}}$ (renormalization of $\sim 45$) at low temperatures, making it a moderate heavy fermion system \cite{das14a,das15a}. PrCrGe$_3$, on the other hand, exhibits a smaller $\gamma$ of about $66\,{\rm mJ\,mol^{-1}K^{-2}}$ but shows a large nuclear specific heat contribution ($\alpha_{\rm n}$) at low temperatures. As shown in figure \ref{f2a}, below 1~K, $C(T)/T$ for PrCrGe$_3$ increases dramatically due to the nuclear Schottky effect which is caused by the strong interaction between the nuclear magnetic moments with the strong magnetic field produced by the 4$f$ electrons at the nuclear site resulting in the splitting of the nuclear hyperfine levels \cite{grivei95a,lounasmaa62a,steppke10a}. 
Low temperature $C(T<2\,{\rm K})$ data  of PrCrGe$_3$ were fitted with $C/T= \gamma + \alpha_{\rm n} T^{-3}.$
Remarkably, $\alpha_{\rm n}$ amounts to $\rm{\sim454\,mJ\,K\,mol^{-1}}$ which is significantly higher than $\alpha_{\rm n}$ of pure Pr metal ($\rm{\sim56\,mJ\,K\,mol^{-1}}$) at zero applied field \cite{pathak13a} suggesting the presence of a strong hyperfine interaction in PrCrGe$_3$. 

\subsection{Electron Spin Resonance}

\begin{figure}[h]
 \centering
\includegraphics[width=0.55\textwidth]{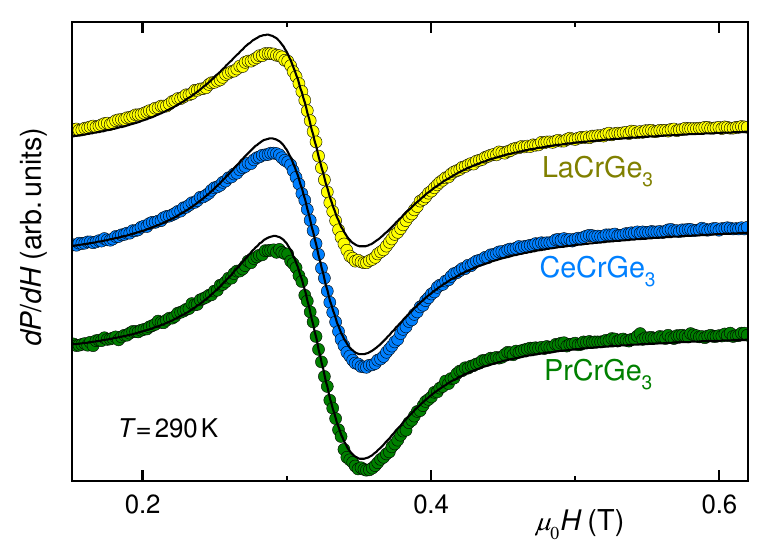} 
\caption{ESR spectra (symbols) at $T=290$~K for LaCrGe$_{3}$, CeCrGe$_{3}$, and PrCrGe$_{3}$ and Lorentzian shape (lines) with $g=2.10\pm0.02$ and $\mu_{0}\Delta H = 55\pm2$~mT.}
\label{signals}
\end{figure}

Typical ESR spectra at room temperature, in the paramagnetic state of LaCrGe$_{3}$, CeCrGe$_{3}$, and PrCrGe$_{3}$ are shown in figure \ref{signals}. For all compounds the spectra could be well described by symmetric Lorentzian line shapes, indicating sample grain sizes being smaller than the microwave penetration depth. This resulted in the powder-averaged parameters $g=2.10\pm0.02$ and $\mu_{0}\Delta H = 55\pm2$~mT. The deviations from the Lorentzian shapes are due to an anisotropy of the $g$ factor which, however, could not be reliably resolved by fitting with powder-averaged Lorentzians due to broad lines and broad background structures.

The obtained $g$ factor describes the $S=3/2$ spins of Cr$^{3+}$ ions which are centered in octahedra which form face sharing chains in the hexagonal (P6$_{3}/mmc$ space group) crystal structure. However, for an insulating environment and in an octahedral field, one expects for local Cr$^{3+}$ (Hund's rule $^{4}F_{3/2}$ state) a $g$ value slightly below 2 for the lowest lying level which is effectively an $S$ state because of the quenched orbital moment \cite{abragam70a}.
%
\begin{figure}[hbt]
 \centering
\includegraphics[width=0.37\textwidth]{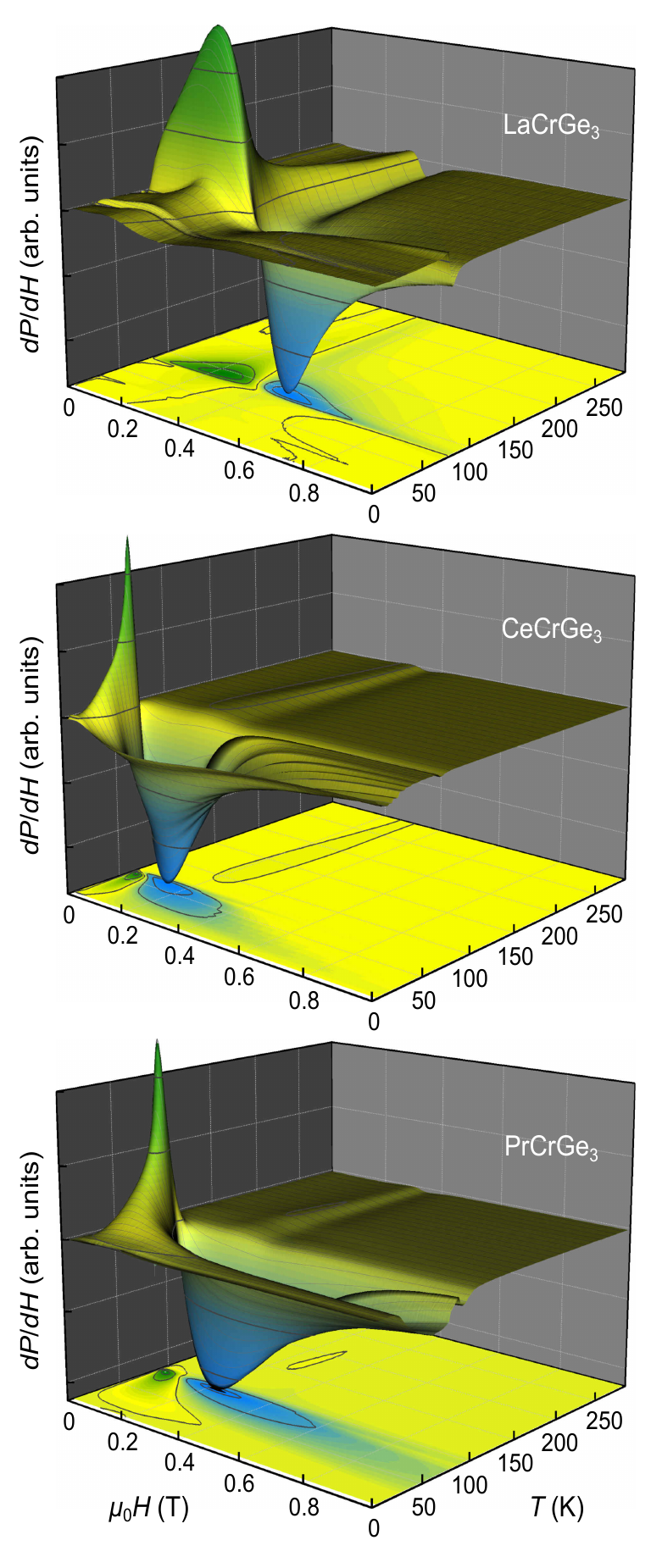}
\caption{Temperature dependence of the ESR spectra $dP/dH (H)$ for LaCrGe$_{3}$, CeCrGe$_{3}$ and PrCrGe$_{3}$.}
\label{ESRSpecTemp}
\end{figure}
%
In order to explain the deviation from the observed $g$ value, $\Delta g$, one needs to take into account that the Cr 3$d$ states are delocalized in narrow $d$-bands, leading to the band ferromagnetism in all three compounds \cite{bie07a}. Therefore, the resonance may be interpreted as a conduction electron spin resonance of conduction spins in a narrow band which is based almost entirely on Cr $d$-states. Then $\Delta g$ is an effect of spin-orbit coupling of the conduction electrons and depends on the band structure \cite{dora09a}.

Figure \ref{ESRSpecTemp} illustrates the temperature dependence of the derivative microwave absorption for the three compounds. For temperatures above the ordering temperatures the wavy features correspond to the paramagnetic resonance signal. Interestingly, in LaCrGe$_{3}$ this feature is also visible in the ordered temperature region indicating a ferromagnetic resonance signal. Close to the magnetic phase transitions the microwave absorption is dominated at low fields ($<0.1$~T) by pronounced non-resonant structures reflecting a strongly field dependent magnetization. These structures have been disregarded when fitting the derivative microwave absorption with a Lorentzian shape.
\begin{figure*}[hbt]
 \centering
\includegraphics[width=0.95\textwidth]{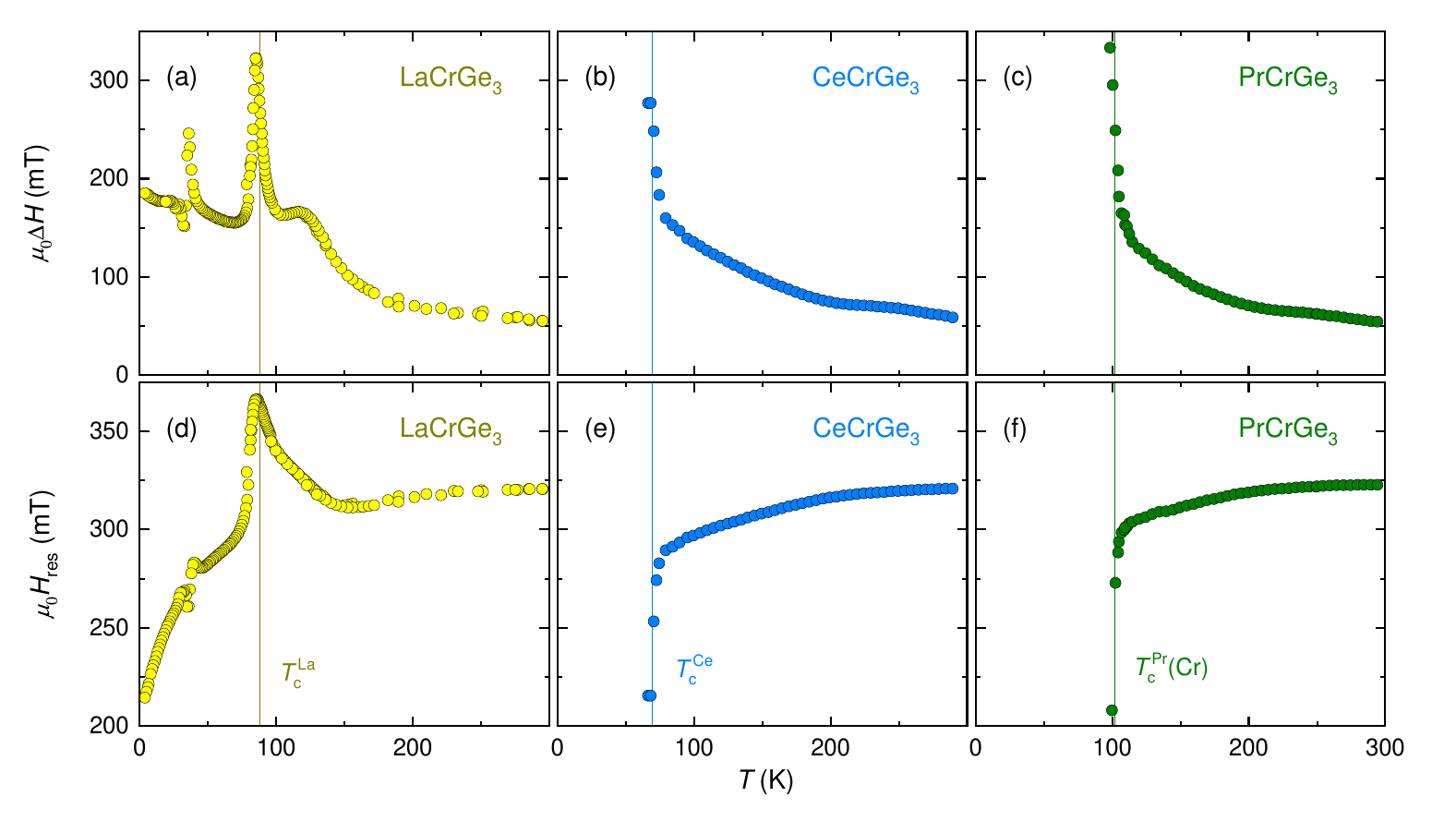}
 \caption{Linewidth $\mu_{0}\Delta H$ and resonance field $\mu_{0}H_{\rm res}$. $T_{\rm C}$ indicates the ferromagnetic ordering temperature. $T_{\rm C}$ was determined by the point of inflection in the $\chi(T)$ data.}
\label{ESRparam}
\end{figure*}

The results of a Lorentzian line fitting are displayed in figure \ref{ESRparam}. The effect of ferromagnetic ordering is to shift the line towards lower resonance fields which is nicely seen for LaCrGe$_{3}$. For CeCrGe$_{3}$ and PrCrGe$_{3}$ the line becomes undetectable due to extreme broadening and shift. The reason for this quite different behavior to LaCrGe$_{3}$ may be the much stronger magnetic anisotropy in CeCrGe$_{3}$ and PrCrGe$_{3}$ which arises from the 4$f$ states.  
In the ordered state of LaCrGe$_{3}$ demagnetization fields could be a relevant contribution to the observed resonance field. 
Assuming for the investigated polycrystalline powder spherically shaped samples with a demagnetizing factor of $\mathcal{N}=1/3$ we obtain demagnetization $\mathcal{N}M(T,\mu_0H=320$\,mT) values below 0.1\,mT in the entire temperature range. 
Thus, this ``sample-shape anisotropy`` has a tiny effect, leaving anisotropy fields as a major source for the observed line shift.

\section{Discussion and Conclusion}

The results of the ESR study on LaCrGe$_{3}$, CeCrGe$_{3}$ and PrCrGe$_{3}$ are similar in the paramagnetic region but strongly depend on the presence of 4$f$ electrons for temperatures close to and below the ferromagnetic ordering temperature $T_{\rm C}$, see figure \ref{ESRparam}. 
The coupling between 4$f$ and 3$d$ magnetism obviously influences the spin resonance of Cr 3$d$ electrons. Therefore, the effect of 4$f$ magnetism seems to be relevant at relatively high temperatures although 4$f$ magnetic ordering occurs well below the ordering of the 3$d$ moments. For example, in PrCrGe$_{3}$ the Pr ordering is reflected in an additional increase of the magnetic susceptibility below 30~K (see figure \ref{props}a) whereas 3$d$ magnetic ordering is at $T_{\rm C}=102$~K.
 Approaching $T_{\rm C}$ in the paramagnetic region, the linewidth increases due to critical fluctuations and a reduction of exchange narrowing processes \cite{anderson53a} whereas the resonance field decreases because internal fields are built up towards $T_{\rm C}$. Below $T_{\rm C}$, in PrCrGe$_{3}$ and CeCrGe$_{3}$, the anisotropy inherent in the 4$f$ magnetism leads to additional broadening such that the 3$d$ resonance is unobservable below $T_{\rm C}$ (see figures \ref{ESRparam}b, c, e and f).   

In case of LaCrGe$_{3}$ (4$f^{0}$), the pure 3$d$ magnetism determines the resonance properties. Besides the clear signatures at ferromagnetic ordering there are also anomalies around 35~K and 125~K in the linewidth and resonance field (see figures \ref{ESRparam}a and d, respectively).
Around 35~K, measurements of the thermoelectric power show a broad hump which was related to the ordering of Cr moments \cite{das14a}.
The small maximum in the linewidth data around 125~K has no correspondence to other quantities and is the result of a broad, unresolved background line superimposed on the main line. Upon decreasing the temperature below 125~K the main line starts to dominate and the fitted ESR parameters approach the main line properties.

Interestingly, there is no clear difference in the temperature dependence of the ESR parameters between PrCrGe$_{3}$ and CeCrGe$_{3}$ compounds.  
However, one would expect a difference since both compounds show a different coupling among their 4$f$-3$d$ electron systems. PrCrGe$_{3}$ (4$f^{2}$) has a stable, classic 4$f$ moment [a non-magnetic singlet is possible which, however, is not indicated in the magnetic susceptibility (see upturn in $\chi(T)$ below 30~K, figure \ref{props}a)]. CeCrGe$_{3}$ (4$f^{1}$) is a heavy fermion ferromagnet where 4$f$ and conduction electron (3$d$) states are hybridized leading to Kondo lattice behavior of the electrical resistivity and an enhanced Sommerfeld coefficient of the specific heat \cite{das14a}. The Kondo effect was shown to be essential for the observability of a Yb$^{3+}$-based heavy electron spin resonance in YbRh$_{2}$Si$_{2}$ \cite{kochelaev09a,wolfle09a}. In this respect, for temperatures above $T_{\rm C}$, it seems remarkable that the Cr 3$d$ resonance is very similar in PrCrGe$_{3}$ and CeCrGe$_{3}$, regardless of the presence or absence of 4$f$ conduction-electron hybridization.   

\section*{Acknowledgements}
We acknowledge valuable discussions with Christoph Geibel. ZH would like to acknowledge the Polish National Agency for Academic Exchange (NAWA) for ULAM fellowship.

\section*{References}

\providecommand{\newblock}{}

\end{document}